\documentclass[twocolumn]{iopart}

\usepackage{eurosym}
\usepackage[utf8]{inputenc}
\usepackage{amsfonts}
\usepackage{amssymb}
\usepackage{graphicx}
\usepackage{pgfplots}
\usepackage{geometry}
\usepackage{titling}
\usepackage{cite}
\usepackage{tikz}
\usepackage{geometry}
\usepackage{float}
\usepackage{caption}
\usepackage{multicol}
\usepackage{xcolor}
\usepackage[Glenn]{fncychap}
\usepackage{subcaption}
\usepackage{array}
\usepackage{booktabs}
\usepackage[colorlinks=true, linkcolor=blue, citecolor=blue, filecolor=magenta, urlcolor=blue]{hyperref}

\usepackage{iopams}  

\usepackage{titlesec}
\titleformat{\section}{\Large\bfseries}{\thesection}{1em}{}
\titleformat{\subsection}{\large\bfseries}{\thesubsection}{1em}{}

\begin{document}
\onecolumn

\title{Magnetic Phase Transitions and Mixed Spin\\in Double Perovskite $Sr_{2}FeMoO_{6}$}

\author{KHAIREDDINE Said{\textsuperscript{1}} - ASSAD Redouane{\textsuperscript{1}}{\textsuperscript{3}} \\ EL FALAKI Mohammed{\textsuperscript{4}} - AHL LAMARA Rachid{\textsuperscript{1}} - DRISSI Lalla Btissam{\textsuperscript{1}{\textsuperscript{2}}}}

\address{ {\textsuperscript{1}} LPHE-MS, Faculty of Science, Mohammed V University in Rabat, Morocco.
\\ {\textsuperscript{2}} CPM, Centre of Physics and Mathematics, Faculty of Science, \\Mohammed V University in Rabat, Morocco. \\ {\textsuperscript{3}} National Center for Scientific and Technical Research (CNRST), Morocco.\\ {\textsuperscript{4}} Laboratory of Innovation in Science, Technology and Modeling, Faculty of Sciences, Chouaib Doukali University, El Jadida, Morocco}

\ead{said\_khaireddine@um5.ac.ma}
\vspace{10pt}
\begin{indented}
\item[]March 2025 

\end{indented}
\begin{abstract} The magnetic properties of the double perovskite oxide $Sr_{2}$FeMo$O_{6}$ are analyzed using a mixed-spin Ising model with spins $\left( \frac{1}{2},\frac{5}{2}\right) $ in the presence of a random crystal field \(\Delta\) and exchange interactions $ J $ on a three-dimensional (3D) cubic lattice. The study employs both the Mean-Field Approximation (MFA) based on the Bogoliubov inequality for Gibbs free energy and Monte Carlo (MC) simulations using the Metropolis algorithm to provide a comprehensive analysis of the system's phase transitions and magnetization behavior, with focusing on the role of Fe and Mo sublattices. We establish the ground-state phase diagram, identifying multiple stable magnetic configurations and first-order transitions at low temperatures. Indicate compensation temperature $T_{comp}$. This work provides deeper insight into the thermodynamic, physics statistic and magnetic properties of $Sr_{2}$FeMo$O_{6}$, with implications for future applications in spintronics and magnetic storage technologies.

\vspace{1pc}
\end{abstract}

\twocolumn
\setlength{\mathindent}{0pt}
\section{Introduction}

The exploration of magnetic materials has witnessed remarkable advancements,
underpinning innovations in data storage, quantum computing, and
spintronics. Recent studies have underscored the unique properties of
two-dimensional (2D) and three-dimensional (3D) magnetic systems, which
exhibit distinct magnetic behaviors due to their reduced dimensionality \cite%
{A2,A3,A4}. 2D magnetic materials have revealed phenomena like long-range
ferromagnetism at the atomic layer scale, offering exciting prospects for
nanoscale devices \cite{A5, A6}. On the other hand, 3D magnetic materials
remain vital for applications requiring bulk properties, including high
magnetic anisotropy and stability. The ability to manipulate spin
interactions, magnetic ordering and phase transitions, continues to inspire
extensive theoretical and experimental investigations \cite{A006,A06}.
Innovations in fabrication techniques have enabled the creation of
high-quality magnetic materials, facilitating a deeper understanding of
their fundamental properties and potential applications in next-generation
devices \cite{1A7,A7}.

Perovskites, a class of materials characterized by their ABX$_{3}$ crystal
structure, exhibit a wide range of magnetic phenomena, making them essential
for both theoretical studies and practical applications \cite{108}. Recent
literature highlights the significance of perovskites in understanding
complex magnetic interactions, particularly through the study of their
electronic structures and phase transitions \cite{A8, A9}. The versatility
of perovskites allows for the tuning of their magnetic properties via
compositional changes or external stimuli, leading to enhanced performance
in applications such as magnetic sensors and memory devices \cite{AA1,AA2,AA3}. The interplay between structural characteristics and magnetic
behavior in perovskite materials is strongly influenced by factors such as
cation ordering, oxygen vacancies, and lattice distortions. For example, the
symmetry and distortion of the perovskite lattice directly impact
spin-polarized electronic transport, critical for advanced functionalities
in spintronic devices \cite{A27}.
Double perovskites represent a fascinating subset of perovskite materials,
where the non-identical cations' arrangement within the crystal lattice,
leads to intricate magnetic interactions that can be explored both
theoretically and experimentally \cite{7, A11}. Recent studies have
demonstrated that double perovskites can exhibit a rich variety of magnetic
orders, including ferro/ferrimagnetism and antiferromagnetism, depending on
their composition and structural configuration \cite{A12, A16}. These mixed
alloys allowed overcoming several challenges related to spin generation,
long distance spin transport, and manipulation and detection of spin
orientation \cite{A14}. These results were used to develop spintronic
materials with high spin polarization, and subsequently to improve the
information storage capacity on modern magnetic media and reduce the data
access time \cite{4,6}. In addition to their high data processing speed,
their storage capacity and their advantage cost, magnetic memory devices
have non-volatile data storage and are less affected by the heat problem,
since spin-polarized currents are associated with processing spin moments
rather than moving electrons, resulting in lower Joule heat dissipation \cite%
{5, A16}.

Several theoretical models have been developed to provide insights into the
underlying mechanisms governing magnetism in double perovskites and related materials. Monte Carlo
simulations have emerged as a powerful tool for investigating the magnetic
properties of complex materials. Phase transitions and critical phenomena
have been explored under various conditions \cite{A17}. Recent results from
Monte Carlo studies have revealed exciting insights into the
temperature-dependent behavior of magnetic systems, highlighting phenomena
such as spin reorientation transitions and critical slowing down \cite%
{A18,A19}. In order to better understand the thermodynamic and magnetic
behavior in equilibrium of the mixed systems, several Ising-based
theoretical methods have been developed such as Effective-field theory, Mean
field approximation, Cluster approach \cite{9,16}. On the other hand, a
number of the experimental magnetic properties of the molecular-based
magnetic materials has been numerically reproduced and explained \cite{17,17A}.

Double perovskite oxides are synthesized through methods like solid-state
reaction and sol-gel processes, with the latter offering finer particle
sizes and better homogeneity, approximately $0.6 \mu m$, compared to 
$0.9\,\mu m$ for the solid-state method \cite{A22}. These materials
exhibit high Curie temperatures and magnetoresistance, with advanced
techniques like spark plasma sintering (with grain sizes ranging from $A%
0.9\,\mu m$ to $1.3\,\mu m$) further enhancing their
properties for multifunctional electronic applications \cite{A20, A21}. The
study of magnetic properties in materials such as Sr$_{2}$FeMoO$_{6}$ has
garnered significant interest due to their complex interactions and
potential applications in spintronics and magnetic storage devices. The half
metallicity of Sr$_{2}$FeMoO$_{6}$ is the physical origin of its full spin
polarization at the Fermi level and well above room temperature, which makes
double perovskite oxide Sr$_{2}$FeMoO$_{6}$ an appropriate material for
spintronic technology \cite{8}. Sol-gel synthesis produces finer powders
with better properties, as $Sr_{2}$FeMo$O_{6}$ exhibits high Curie
temperature ($T_c =420\,K$) and significant saturation magnetization
(up to 18 $\mu$/g) \cite{A22, A24}. Anti-site disorder and oxygenvacancies lower saturation magnetization below the theoretical value of $4\,\mu _{B}$ \cite{A23}. Furthermore, B-site cation ordering significantly
enhances magnetic interactions and Curie temperatures by stabilizing the
ferrimagnetic state in Sr$_{2}$FeMoO$_{6}$ \cite{A28, A29}. While its synthesis and applications focus on achieving high purity and precise
stoichiometry, large gaps remain in understanding the effects of random
crystal fields and anti-site disorder \cite{A26, A25}.

In this paper, we study the magnetic properties of Sr$_{2}$FeMoO$_{6}$ using
a mixed spins ($\frac12$; $\frac52$) Ising model in the presence of the random crystal
field in a three-dimensional cubic lattice. First, we will study the
magnetic behavior of the system within the framework of the mean field
approximation based on the Bogoliubov inequality for Gibbs free energy, then the method of Monte Carlo simulation.
We will study magnetic anisotropy effects on the magnetization of the two
sublattices with spins $\frac12$and $\frac52$ for different values of the temperature and the exchange
interaction. We also present the effect of the exchange interaction and the
magnetic anisotropy on the critical temperature.

\section{Model and formalism}

To study Sr$_{2}$FeMoO$_{6}$, we begin   by analyzing its crystal structure, focusing on the Fe and Mo cation ordering and its impact on symmetry. The electronic structure is explored through Fe and Mo d-orbitals and spin polarization.
In the second part, a mixed-spin Ising model is applied using two methods: the mean-field approximation and Monte Carlo simulations.

\subsection{Sr$_{2}$FeMoO$_{6}$ Structure and Magnetic moments}

\begin{figure}[]
\centering
\includegraphics[width=0.45\textwidth]{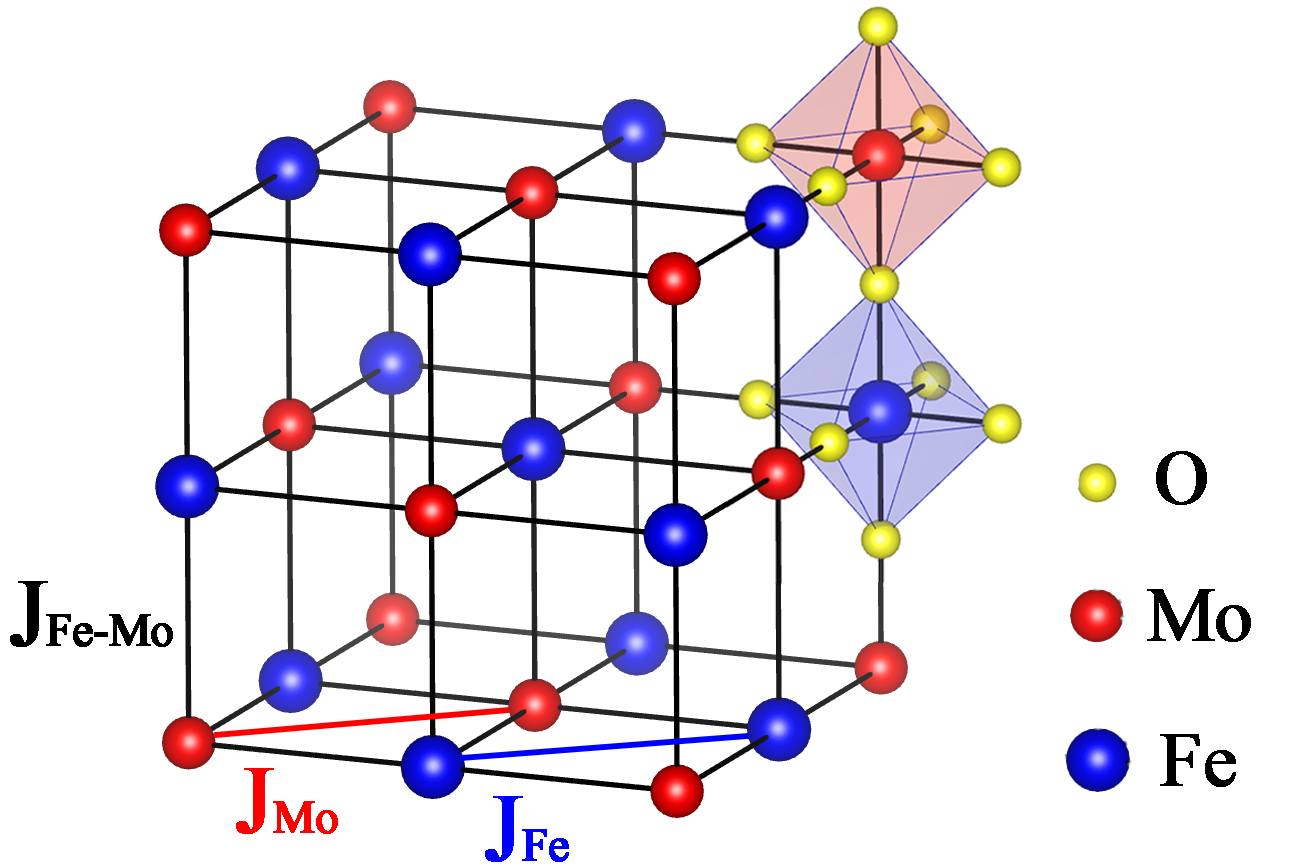}
\caption{{\protect\small The ideal structure of a double perovskite Sr$_2$FeMoO$_6$. The Sr$%
^{2+}$ cations, located in the center of each cell, are not represented.}}
\label{fig:1}
\end{figure}

Double perovskite oxide Sr$_{2}$FeMoO$_{6}$ has a cubic or tetragonal structure
depending on the Fe and Mo atoms ordering and stoichiometry \cite{18}. In
the cubic ideal structure Sr$_{2}$FeMoO$_{6}$, iron (Fe) and molybdenum (Mo)
cations are inside an oxygen octahedron, forming the octahedra FeO$_6$ and
MoO$_6$ attached by the corners. The alkaline earth metal cations
(strontium) are positioned body-centered into the octahedral sites \cite{8}.
The ideal structure of Sr$_{2}$FeMoO$_{6}$ composition, which can be affected by
the disorder of the cations Fe and Mo in the lattice, is represented by an
orderly arrangement and a three-dimensional alternation: each octahedron FeO$%
_6$ has as neighbors only MoO$_6$ octahedra and vice versa (see  \hyperref[fig:1]{Figure \ref{fig:1}}).

The Fe$^{3+}$ cation is in its fundamental energetic state with 5 electrons in 3d orbit, which generates a relatively localized $\frac52$ moment (comparatively to the highly localized moment of the Sr$^{2+}$ cation),
while the total electronic density of states for the molybdenum cation Mo$%
^{5+}$ is essentially determined by a single electron in orbit 4d$^1$ forming a delocalized $\frac12$
moment. The 4d$^1$ electrons of the Mo$^{5+}$ are itinerant and form a
conduction band. Thereby, magnetic properties of Sr$_2$FeMoO$_6$ are highly
sensitive to the Fe/Mo ratio content.

\begin{figure}[]
\centering
\includegraphics[width=0.4\textwidth]{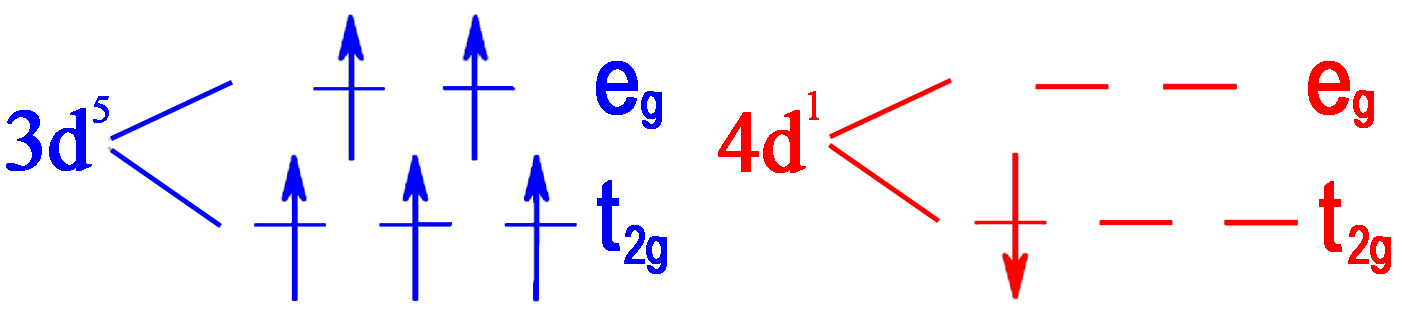}
\caption{{\protect\small Electronic Configurations of cations Fe$^{3+}$ (3d$^5$ : spin up $\frac52$) and Mo$^{5+}$ (4d$^1$ : spin down $\frac12$) $S_{total} = 2$}}
\label{fig:2}
\end{figure}

Hence, the ferrimagnetic half metallic state of Sr$_2$FeMoO$_6$ is due to a
large antiferromagnetic superexchange interaction between the $\frac52$
spins of Fe$^{3+}$ and the $\frac12$ spins of Mo$^{5+}$. Electrons transfer is only possible if iron spins are all oriented antiparallel with respect to the spins of the itinerant electrons of Mo$^{5+}$ because of the Pauli exclusion principle (see \hyperref[fig:2]{Figure \ref{fig:2}}) .

The antiferromagnetic coupling of the sublattices has been demonstrated by
X-ray circular magnetic dichroism \cite{19}. The consequence of such a type
of coupling will be a large saturation moment $4\mu_B$ (equivalent to $S_{total} = 2$). This theoretical value is never reached and the
measurements show the existence of a weaker magnetization and always less
than $3.8\mu_B$.

\begin{figure}[]
\centering
\includegraphics[width=0.45\textwidth]{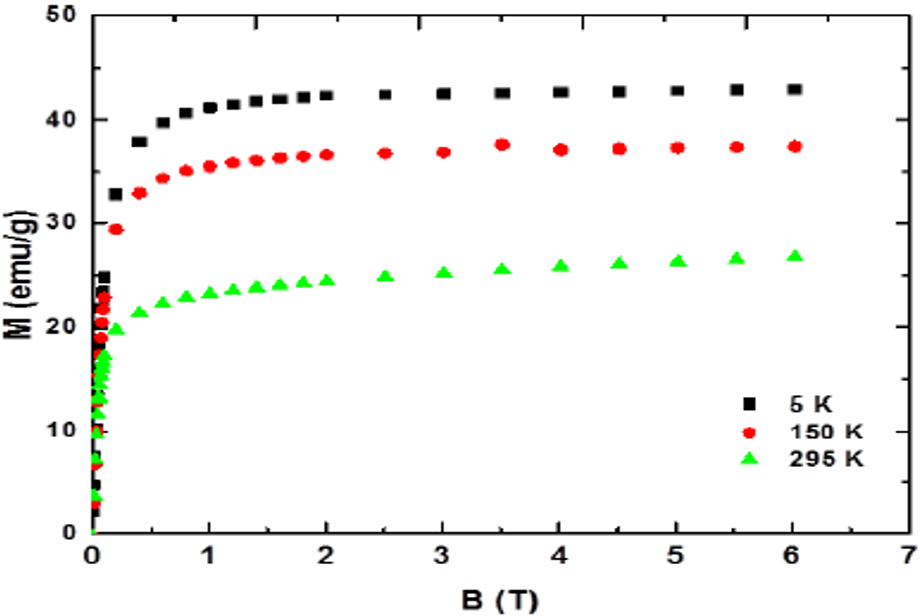}
\caption{{\protect\small Magnetization curves on Sr$_2$FeMoO$_6$ sample for three different
temperatures \protect\cite{20}.}}
\label{fig:3}
\end{figure}

Additionally, the measurements of the magnetization as a function of the
external magnetic field at low temperature reveal that the double perovskite
oxide Sr$_{2}$FeMoO$_{6}$ can be successfully used in various applications
because it can be processed in normal conditions (meaning low temperatures
and short times of sintering) \cite{20}. In fact, at three low different
temperatures (5 K, 150 K, and 295 K lower than the Currie temperature) the
magnetization increases rapidly at low values of applied magnetic fields,
induced by the magnetic moments alignment of Fe and Mo ions (see \hyperref[fig:3]%
{Figure \ref{fig:3}}).

\subsection{Computational details }

We study the system using Mean-Field Approximation and Monte-Carlo simulations within statistical physics. MFA provides an analytical approach but neglects fluctuations, while MC captures thermal fluctuations with higher accuracy. Through mathematical developments, we combine both methods to analyze phase transitions and critical behavior.

Here, we consider a mixed-spin Ising system to model double perovskite Sr$%
_{2}$FeMoO$_{6}.$ In this model, consisting of two sublattices (I, II),
interactions occur between the $S=\frac52$ spins of Fe$^{3+}$ and the $\sigma
=\frac12$
spins of Mo$^{5+}$, and a crystal field $\Delta $ is mainly acting on spins $%
S=\frac{5}{2}.$ In this case, $\sigma{i}$ spins with values $\pm \frac{1}{2}$
occupy the first sublattice (I) and $S_{j}$ spins with six values $\pm \frac{%
5}{2}$, $\pm \frac{3}{2}$ and $\pm \frac{1}{2}$ will occupy the second
sublattice (II). In the case of $N$-atoms, the Ising's Hamiltonian of this
system is given by:

\begin{equation}
        \begin{array}{ll}
        H= & -J_{Fe}\sum_{(i,j)}^{\frac{N}{2}}S_{i}S_{j}-J_{Mo}\sum_{(i,j)}^{\frac{N%
        }{2}}\sigma _{i}\sigma _{j} \\ 
        & -J_{Fe-Mo}\sum_{(i,j)}^{N}S_{i}\sigma _{j}-\Delta \sum_{i=1}^{\frac{N}{2}}S_{i}^2 \\
        & -h\sum_{i=1}^{\frac{N}{2}}(S_{i}+\sigma _{j})%
        \end{array}  \label{eq1}
\end{equation}

where $J_{Fe}$, $J_{Mo}$, and $J_{Fe-Mo}$ are exchange
interactions between near-neighbor pairs of spins $S-S$, $\sigma-\sigma$, and $S-\sigma$
respectively, and $\Delta $ is the crystal field of spin $S_{Fe}$. Each spin is surrounded by 6 spins first nearest neighbors and 12 spins as second
near neighbors. In order to reduce the number of parameters and simplify the analysis,  all our results will be normalized by the coefficient exchange $J_{Mo}$.

\subsection{Mean-Field Approximation}

The Mean-Field Approximation is a fundamental method in statistical physics that simplifies complex interactions by averaging them, providing an analytical approach to phase transitions. However, it neglects thermal fluctuations, making it less accurate near critical points. The variational principle based on the Gibbs-Bogolubov inequality for the free energy per site \cite{25}.\newline
The effective Hamiltonian corresponding to the mean field,for $S_{i}=S$
and $\sigma =\sigma _{i},$ is written as:
\begin{equation}
H_{0}=h_{I}\sum_{i}^{\frac{N}{2}}S_{i}+h_{II}\sum_{i}^{\frac{N}{2}%
}\sigma_{i}-\Delta \sum_{i}^{\frac{N}{2}}S_{i}^{2}
\end{equation}%
with

\begin{equation}
\begin{array}{rl}
 h_I = -z_2 J_{Fe} m_{Fe} - z_1 J_{Fe-Mo} m_{Mo} - h \\
h_{II} = -z_2 J_{Mo} m_{Mo} - z_1 J_{Fe-Mo} m_{Fe} - h 
\end{array}
\end{equation}

Here, $z_{1} = 6$ and $z_{2} = 12$ are the nearest neighbors and the second near
neighbors respectively.
\newline
According to the variational principle, the free energy is expressed as follows:

 the inequality
is given by: 
\begin{equation}
F\leq F_{0}=-\frac{K_{B}T}{N}\ln (Z)+\frac{1}{N}\langle H-H_{0}\rangle _{0}
\end{equation}
where the term $F_{0}$ is defined as:%

\begin{equation}
\begin{array}{rl}
 F_{0}=&-\frac{K_{B}T}{N}\ln(Z) -\frac{1}{2}J_{Fe-Mo}z_{1}m_{Fe}m_{Mo}  \\
&-\frac{1}{2}h_{I}m_{Fe} -\frac{1}{2}h_{II}m_{Mo}  -\frac{1}{4}J_{Fe}z_{2}m_{Fe}^{2} \\
&-\frac{1}{4}J_{Mo}z_{2}m_{Mo}^{2}   
\end{array} \label{eq 6}
\end{equation}

given terms of the Boltzmann constant ($K_{B}$ ), that we have fixed
at $K_{B}=1$ for sake of simplicity. $Z$ is the partition function, $H$ is the Hamiltonian of the system, and $H_{0}$ is the effective Hamiltonian.  The partition function generated by the above
Hamiltonian $H$  has the form:

{\small
\begin{equation}
\begin{array}{rl}
Z =&\left[2e^{\frac{25}{4}\beta\Delta}\cosh\left(\frac{5}{2}\beta h_I\right) + 2e^{\frac{3}{4}\beta\Delta}\cosh\left(\frac{3}{2}\beta h_I\right) \right. \\  
& \left. + 2e^{\frac{1}{4}\beta\Delta}\cosh\left(\frac{1}{2}\beta
h_I\right)\right] \times\left[2\cosh\left(\frac{1}{2}\beta h_{II}\right)\right] 
\end{array}
\end{equation}
}

In order to study the magnetization of the system, the sublattice
magnetization per site of the Fe ($m_{Fe}$) and Mo ($m_{Mo}$) are
calculated. These order parameters ($m_{Fe}$ and $m_{Mo}$) are defined by
minimizing the free energy with respect to the effective fields $h_{I}$ and $%
h_{II}$ respectively. They are given by:

\begin{equation}
\begin{array}{rl}
 m_{Fe} &= -\frac{A_{1}}{A_{2}} \\ m_{Mo} &= -\frac{1}{2}\tanh\left(\frac{1}{2}\beta h_{II}\right) 
\end{array}
\label{eq 7}
\end{equation}

with

{\small 
\begin{equation}
\begin{array}{rl}
A_{1} =& \frac{5}{2}e^{\frac{25}{4}\beta\Delta}\sinh\left(\frac{5}{2}\beta h_{I}\right) +\frac{3}{2}e^{\frac{9}{4}\beta\Delta}\sinh\left(\frac{3}{2}\beta h_{I}\right) \\ 
& +\frac{1}{2}e^{\frac{1}{4}\beta\Delta}\sinh\left(\frac{1}{2}\beta h_{I}\right) \end{array}
\end{equation}
}

{\small 
\begin{equation}
\begin{array}{rl}
 A_{2} =& e^{\frac{25}{4}\beta\Delta}\cosh\left(\frac{5}{2}\beta h_{I}\right) + e^{\frac{9}{4}\beta\Delta}\cosh\left(\frac{3}{2}\beta h_{I}\right) \\ 
& + e^{\frac{1}{4}\beta\Delta}\cosh\left(\frac{1}{2}\beta h_{I}\right)
\end{array}
\end{equation}
}

\subsection{Monte Carlo Simulation}

In order to confirm the results previously obtained by Mean Field Approximation, we numerically simulate the Sr$_{2}$FeMoO$_{6}$ magnetic properties based on the Hamiltonian described by \hyperref[eq1]{Eq \ref{eq1}} on a three-dimensional cubic lattice of $N$ spins and volume $L \times L \times L$, using the standard Monte Carlo technique with the Metropolis algorithm \cite{26, 27}. We ran the program for size $L = 4$ and $L = 16$ in order to analyze critical behavior and used $10^4$ Monte Carlo steps to calculate the means of the different physical quantities such as: \newline
The magnetizations by site of the (Fe) and (Mo) sublattices $m_{Fe}$ and $m_{Mo}$:

\begin{equation}
    m_{Fe}=\frac{2}{N}\sum_{i}S_{i}\quad ;\quad m_{Mo}=\frac{2}{N}\sum_{j}\sigma
    _{j}
\end{equation}

The total magnetization per site $M_{T}$: 

\begin{equation}
M_{T} = \frac{m_{Fe} + m_{Mo}}{2}
\end{equation}

Also, the total susceptibility per site $\chi_T$:

\begin{equation}
\chi = \beta N (\langle M_{T}^{2} \rangle - \langle M_{T} \rangle^{2})
\end{equation}

\section{Results and Discussions}

This section explores the magnetic properties and phase transitions in Sr$_{2}$FeMoO$_{6}$. We start with the ground-state phase diagram, identifying stable magnetic configurations at zero temperature. Next, we analyze finite-temperature phase transitions using the mean-field approximation, followed by Monte Carlo simulations.

\subsection{Ground State}

Let's start by analyzing the magnetic behavior of the double perovskite
oxides Sr$_{2}$FeMoO$_{6}$ in the absence of any thermal excitation. Since
the values of the spins $S=\frac{5}{2}$ and $\sigma=\frac{1}{2}$ are known, six different magnetic configurations can be defined:
three ferromagnetic phases corresponding to mixed spins $F_{1}$ $=\left( 
\frac{1}{2},\frac{1}{2}\right) $, $F_{2}$ $=\left( \frac{1}{2},\frac{3}{2}\right) $and $F_{3}$ $=\left( \frac{1}{2},\frac{5}{2}\right) $, two
ferrimagnetic phases resulting from mixed spins $F_{4}$ $=\left( \frac{5}{2}%
,-\frac{1}{2}\right) $ and $F_{5}$ $=\left( \frac{3}{2},-\frac{1}{2}\right) $%
, and finally an antiferromagnetic phase, $F_{6}$ generated by mixed spins $(\frac{1}{2}, -\frac{1}{2})$, We are using free energy to defend stable states. 

The ground-state phase diagram, associated with crystal field $\Delta $,
magnetic field h, and intermediate coupling $J_{Fe-Mo}$, is obtained
exactly from the Hamiltonian of the ground state which is invariant under
translation by comparing the energies of the ground state phases at $T=0K$.
As the results don't depend on the sizes, we limit our calculations to a
fixed double perovskite's size.

By setting, the intra-sublattice couplings $J_{Fe}=J_{Mo}=0.1,$ the
coordination numbers $z_{1}=6$ and $z_{2}=12$, and the external field $h=0,$
the Hamiltonian of Sr$_{2}$FeMoO$_{6}$, given by (\hyperref[eq1]{Eq \ref{eq1}}), can be
simplified to:

\begin{equation}
    H=-\frac{N}{4}\left( (1.2+\Delta )S^{2}+1.2\sigma ^{2}+12J_{Fe-Mo}S\sigma\right)
\end{equation}

To represent $\Delta(J)$, we need to find the intersection points between the energies of different magnetic states, where \(H_i = H_j\). Solving these equations allows us to determine the function $\Delta(J)$ as shown in \hyperref[energy intersections]{Table \ref*{energy intersections}}. Each intersection represents a critical point where the system transitions between different magnetic configurations.

\begin{table}[H]
\centering
\begin{scriptsize}

\begin{tabular}{@{} >{\centering\arraybackslash}m{2.3cm} >{\centering\arraybackslash}m{1.4cm}  >{\centering\arraybackslash}m{2.9cm} @{}}
\toprule
\textbf{Phases } & \textbf{ Magnetic Order} & \textbf{ Coexistence lines } \\ [0.2ex] \hline
\( \left( \frac{5}{2}, \frac{-1}{2} \right) \)  \( \left( \frac{3}{2}, \frac{-1}{2} \right) \) & \( FiM - FiM \) & \(\Delta = 1.5 J_{Fe-Mo} - 1.2\) \\ [0.2ex]
\( \left( \frac{3}{2}, \frac{-1}{2} \right) \)  \( \left( \frac{1}{2}, \frac{-1}{2} \right) \) & \( FiM - AM \) & \(\Delta = 3 J_{Fe-Mo} - 1.2\) \\ [0.2ex]
\( \left( \frac{1}{2}, \frac{1}{2} \right) \)  \( \left( \frac{1}{2}, \frac{-1}{2} \right) \) & \( AM -FM \) & \(J_{Fe-Mo} = 0\) \\ [0.2ex]
\( \left( \frac{1}{2}, \frac{1}{2} \right) \)  \( \left( \frac{3}{2}, \frac{1}{2} \right) \) & \( FM - FM \) & \(\Delta = -3 J_{Fe-Mo} - 1.2\) \\ [0.2ex]
\( \left( \frac{3}{2}, \frac{1}{2} \right) \)  \( \left( \frac{5}{2}, \frac{1}{2} \right) \) & \( FM - FM \) & \(\Delta = -1.5 J_{Fe-Mo} - 1.2\) \\ [0.2ex]
\bottomrule
\end{tabular}
\end{scriptsize}
\caption{{\protect\small Energy formulas and intersection functions for different spin configurations.}}
\label{energy intersections}
\end{table}

\begin{figure}[H]
\centering
\includegraphics[width=0.48\textwidth]{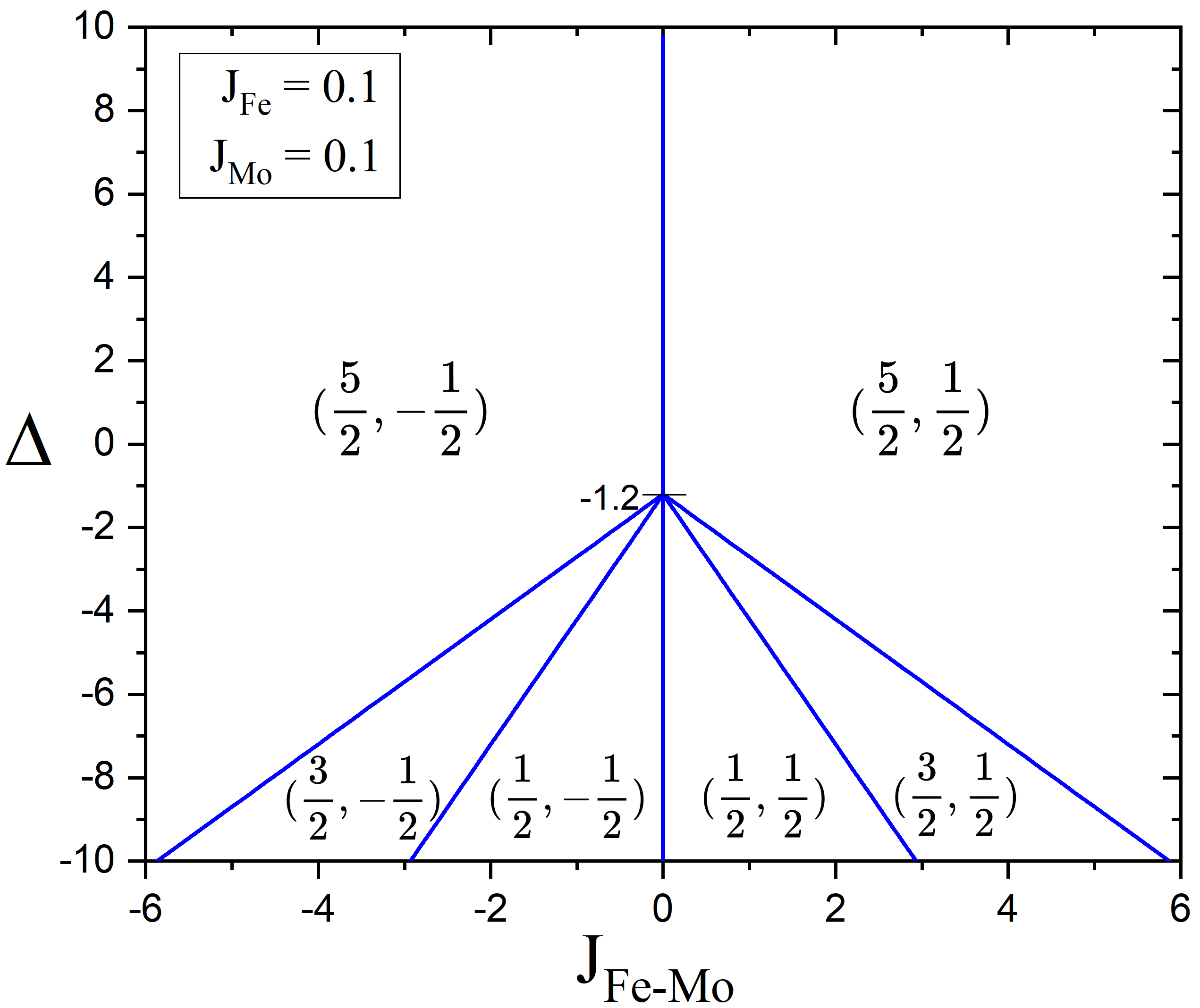}
\caption{{\protect\small Ground state phase diagram $h = 0$ and $J_{Fe} =J_{Mo}=0.1$.}}
\label{fig:4}
\end{figure}

By plotting the functions $\Delta(J)$ for each pair of intersecting magnetic states, we can visualize the ground state on a phase diagram, as illustrated in \hyperref[fig:4]{Figure \ref{fig:4}}. The stable solutions are separated by lines of first-order transition.

As observed in \hyperref[fig:4]{Figure \ref{fig:4}}, two critical regions exist: \(\Delta < -1.2\) and \(\Delta > -1.2\). Consequently, we selected two representative values for \(\Delta\), specifically 5 and -5, as the basis for analyzing our system in the next results.

\subsection{Phase diagrams}

The numeric simulation of the order parameters equations $m_{Fe}$ and 
$m_{Mo}$ \hyperref[eq 7]{Eq \ref{eq 7}} has generated several
solutions, most of them are unstable. The stable solution representing the
state of the system is the one minimizing the free energy $F_{0}$ given in 
\hyperref[eq 6]{Eq \ref{eq 6}}. Transitions are second order if the order
parameters are continuous, and they are first order when the parameters are
discontinuous. The magnetic behavior of the system has been studied through
the order parameter $M_{T}$ (the total magnetization of the system), which
is half the sum of the two order parameters $m_{Fe}$ and $m_{Mo}$. 

\begin{figure}[H]
\centering
\includegraphics[width=0.48\textwidth]{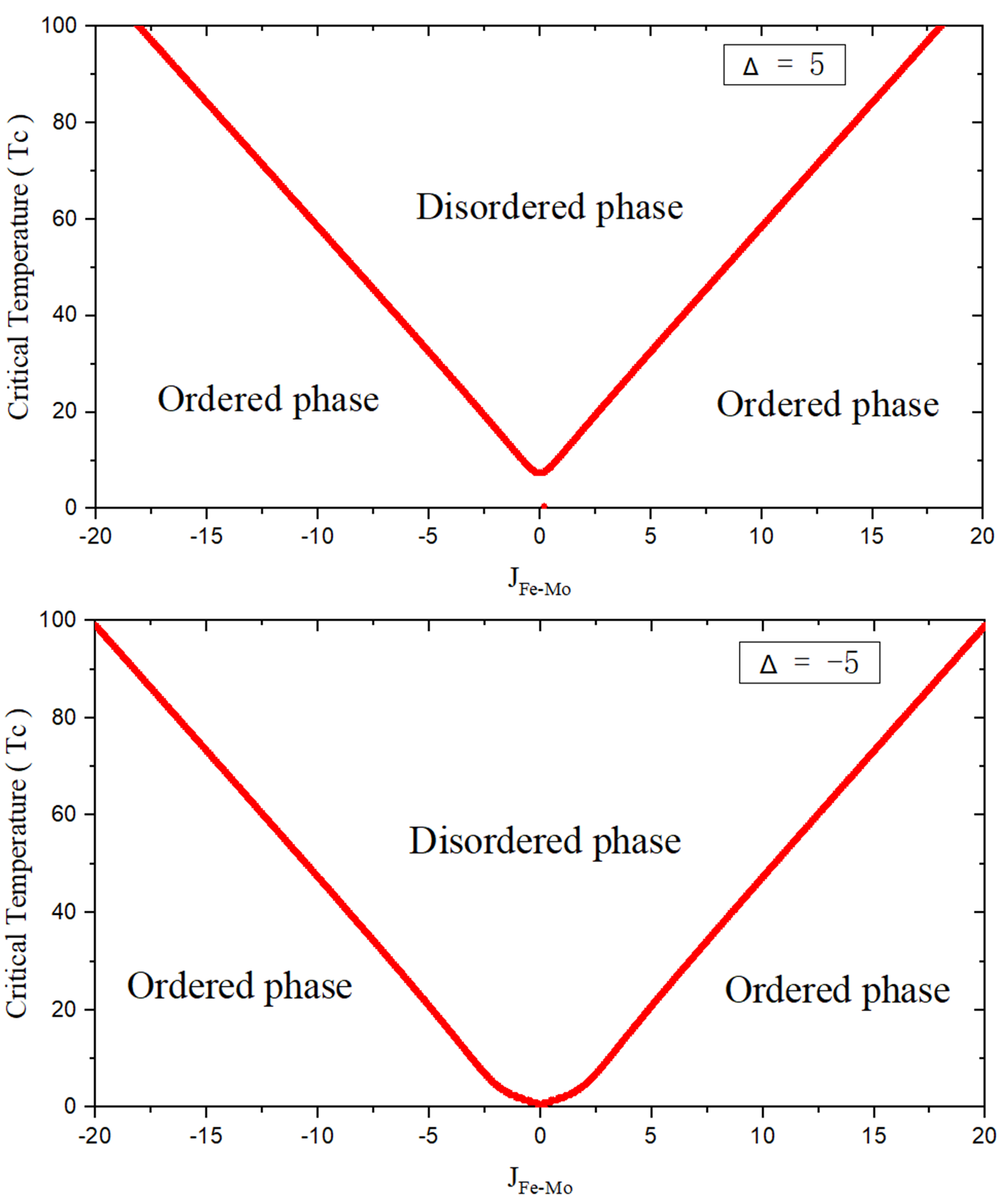}
\caption{The reduced critical temperature $T_c$ as a function of $J_{Fe-Mo}$ for $J_{Fe} = J_{Mo} = 0.1$ and two different values
of $\Delta$ (5 and -5).}
\label{fig:5}
\end{figure}

The reduced exchange coupling $J_{Fe-Mo}$ is an important parameter
in the control of the magnetic order of the system and, therefore, its
critical temperature. We show its influence in the phase diagram plotted in
the $(T_{c},J_{Fe-Mo})$ \hyperref[fig:5]{Figure \ref{fig:5}}, where
we can see that the critical temperature is symmetrical with respect to the
Y axis ($J_{Fe-Mo}$) and can be increased by increasing the reduced exchange coupling in the positive direction by $J_{Fe-Mo}$. As both ferromagnetic and antiferromagnetic couplings can stabilize the ordered phase, provided the magnitude of \( J_{Fe-Mo} \) is large. A stronger exchange coupling (\( |J_{Fe-Mo}| \)) increases the system's thermal stability, raising \( T_c \).

\hyperref[fig:5]{Figure \ref{fig:5}} shows the reduced critical temperature $%
T_c/J$ as a function of the exchange interaction between the two sublattices $J_{Fe-Mo}$ for two values of the magnetic anisotropy ($\Delta$ = 5 and -5 ) Positive $\Delta$ reinforces magnetic order by stabilizing high-spin states, while negative $\Delta$ suppresses it, reducing $ T_c $, In the phase diagram for $\Delta = 5$, the endpoint at $J_{Fe-Mo} = 0.1 $ and $ T_c \approx 0.2 $ marks the critical coupling below which the system transitions from an ordered phase to a disordered phase. This endpoint signifies the limit where sublattice interactions are too weak to sustain magnetic order.

\begin{figure}[H]
\centering
\includegraphics[width=0.48\textwidth]{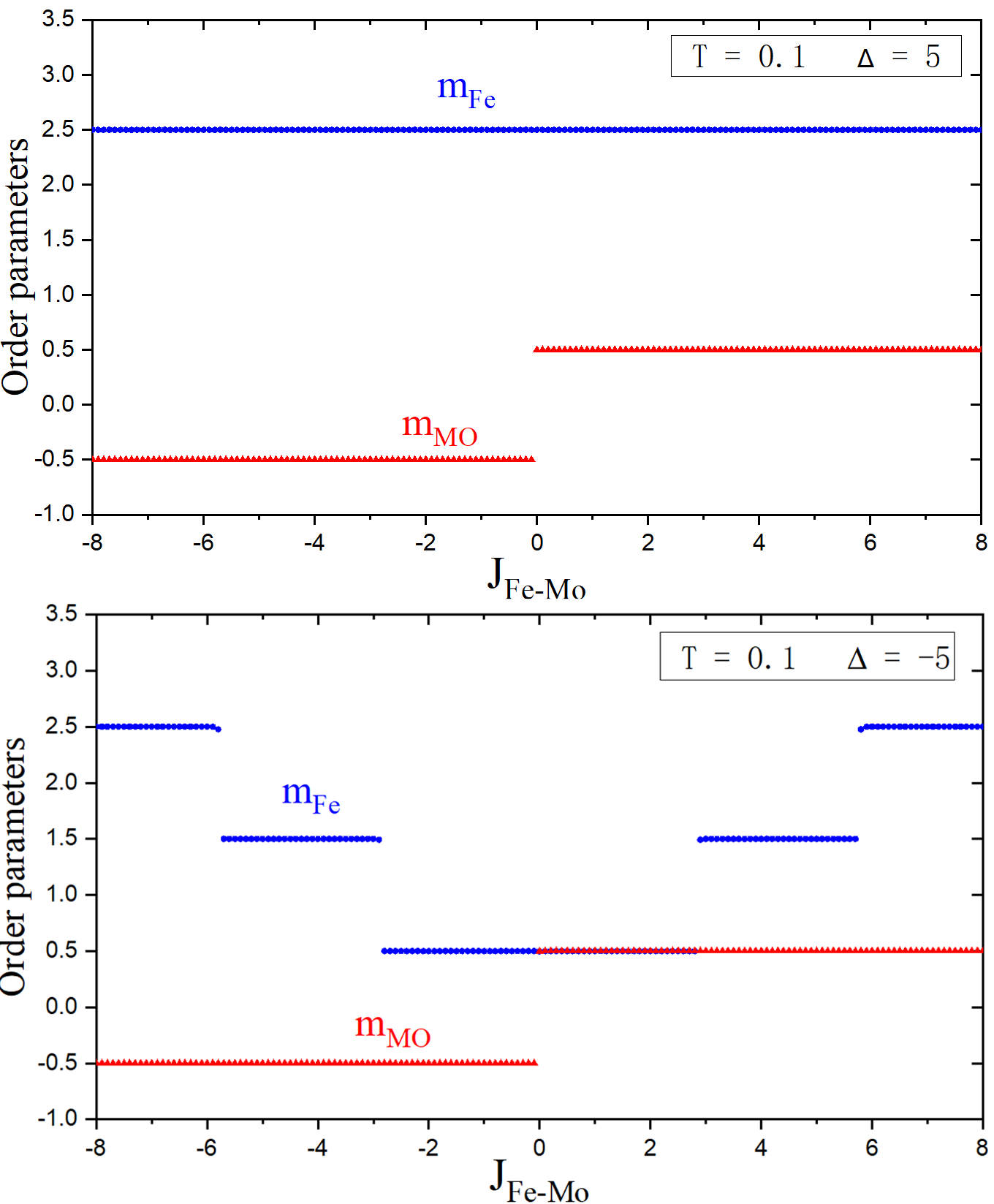}
\caption{Order parameters plotted versus $J_{Fe-Mo}$ for $T = 0.1$
and $J_{Fe} = J_{Mo} = 0.1$ and two values of $\Delta$ (5 and
-5).}
\label{fig:6}
\end{figure}

\hyperref[fig:6]{Figure \ref{fig:6}} illustrates the variation of the order
parameters $ m_{Fe} $ and  $m_{Mo}$ as a function of the coupling constant $J_{Fe-Mo}$ for two different values of the magnetic anisotropy ($\Delta = 5$ and $-5$) at a low temperature (T=0.1) with the same value of the exchange interaction within each sublattice $J_{Fe} = J_{Mo} = 0.1$.

At very low temperatures, the magnetic behavior of the system depends on the
value of the crystal field and the one of the exchange coupling $J_{
Fe-Mo} $. As can be seen in \hyperref[fig:6]{Figure \ref{fig:6}}, the
system changes from a ferrimagnetic phase $\left(\frac{5}{2}, -\frac{1}{2}%
\right) $ to a ferromagnetic phase for positive values of the crystal field $%
\Delta $, However, below a threshold value of the crystal field $\Delta =-5 $ for $J=0.1 $, the system goes from an AFM phase to an FM phase via a succession of first order
transitions, $\left(\frac{5}{2},-\frac{1}{2}\right) \rightarrow \left(\frac{3%
}{2},-\frac{1}{2}\right) \rightarrow \left(\frac{1}{2},-\frac{1}{2}\right)
\rightarrow \left(\frac{1}{2},\frac{1}{2}\right) \rightarrow \left(\frac{3}{2%
},\frac{1}{2}\right) \rightarrow \left(\frac{5}{2},\frac{1}{2}\right) $.
These results are in good agreement with those in the ground state diagram 
\hyperref[fig:4]{Figure \ref{fig:4}}.

\begin{figure}[]
\centering
\includegraphics[width=0.48\textwidth]{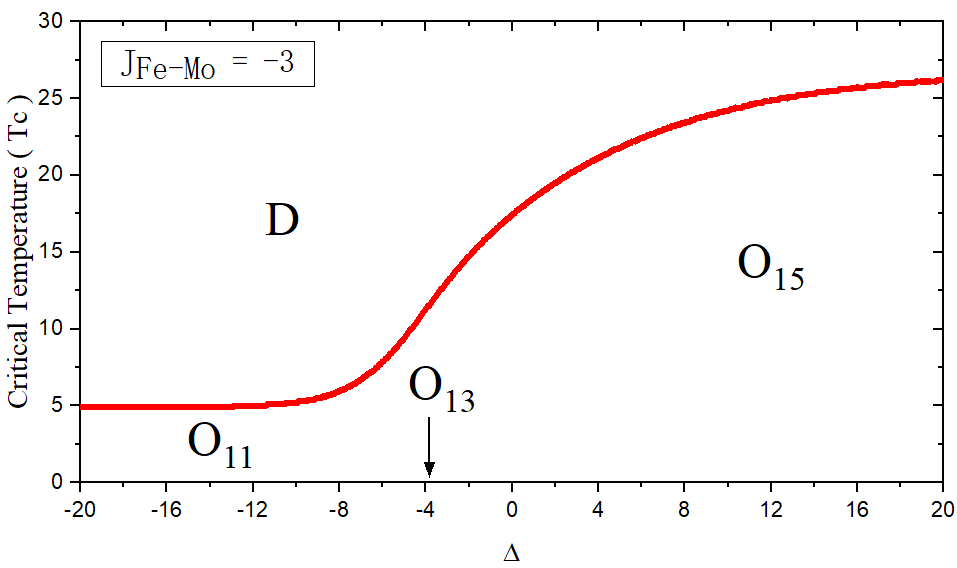}
\caption{The critical temperature $T_c$ as a function of the magnetic
anisotropy $\Delta$ for $J_{Fe-Mo} = -3$ and $J = J_{Fe} = J_{Mo} = 0.1$.}
\label{fig:7}
\end{figure}

The phase diagram presented in \hyperref[fig:7]{Figure \ref{fig:7}} reveals the existence of four distinct phases with different values of the magnetizations by site of the $Fe $ and $Mo $ sublattices $(m_{Fe}, m_{Mo}) $
namely the ordered ferromagnetic phases $O_{11} \equiv (m_{Fe}=\frac{1%
}{2}, m_{Mo}=-\frac{1}{2}) $, $O_{13} \equiv (m_{Fe} =\frac{3}{%
2}, m_{Mo} =-\frac{1}{2}) $ and $O_{15} \equiv (m_{Fe} =\frac{5%
}{2}, m_{Mo}=-\frac{1}{2}) $ and the disordered paramagnetic phase $D
\equiv (m_{Fe}=0, m_{Mo}=0) $.

A second order phase transition line separates the disordered phase D from
the three ordered phases $O_{11} $, $O_{13} $, and $O_{15} $. At very low
temperature, we observe two first order phase transition lines separating
respectively the ordered phases $O_{11} $ and $O_{13} $ and $O_{13} $ and $%
O_{15} $ starting from $(\Delta = -4.94, T = 0) $ and $(\Delta = -2.88, T =
0) $ and ending at two end-points. Above these last two points, regular
sequences occur between each of the two neighboring ordered phases.

\begin{figure}[H]
\centering
\includegraphics[width=0.49\textwidth]{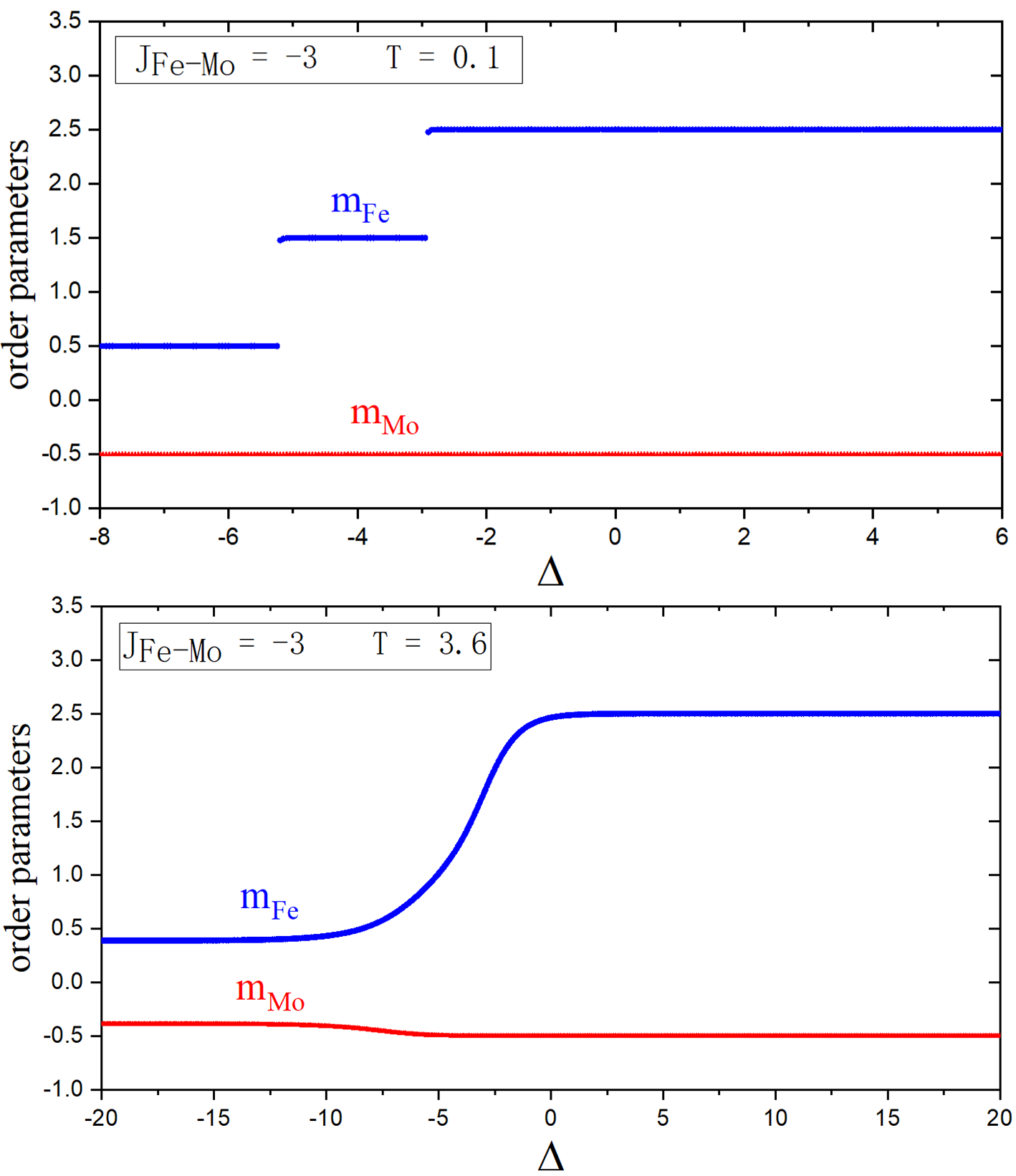}
\caption{Order parameters plotted as a function of the magnetic anisotropy
for $J_{Fe-Mo}=-3$ and $J_{Fe}=J_{Mo}=0.1$ and for two
temperature values (0.1 and 3.6).}
\label{fig:8}
\end{figure}

To show the behavior of the magnetizations of the sublattices $m_{Fe}$
and $m_{Mo}$ as well as the total magnetization $M_T$ in the vicinity
of the first order phase transition lines at very low temperature, we have
studied the variation of these magnetizations as a function of the magnetic
anisotropy ($\Delta$) for $T = 0.1$ and $T = 3.6$ with $J_{Fe} = J_{Mo} = 0.1$ and $J_{Fe-Mo} = -3$ (see \hyperref[fig:8]{Figure %
\ref{fig:8}}). At low temperature ($T=0.1$), the magnetization $m_{Fe}
$ changes discontinuously from the ordered phase $O_{11}$ to the ordered
phase $O_{13}$ at $\Delta= -4.94$ and from the order phase $O_{13}$ to the
ordered phase $O_{15}$ at $\Delta= -2.88$ \hyperref[fig:8]{Figure \ref{fig:8}%
}. These two successive discontinuities correspond to two first order phase
transitions revealed in \hyperref[fig:7]{Figure \ref{fig:7}}. On the other
hand, when the temperature reaches values close to the critical value, the
passage from $O_{11}$ to $O_{13}$ and from $O_{13}$ to $O_{15}$ becomes
continuous as can be seen in \hyperref[fig:8]{Figure \ref{fig:8}}.

\subsection{Temperature effects}

\begin{figure}[H]
\centering
\includegraphics[width=0.48\textwidth]{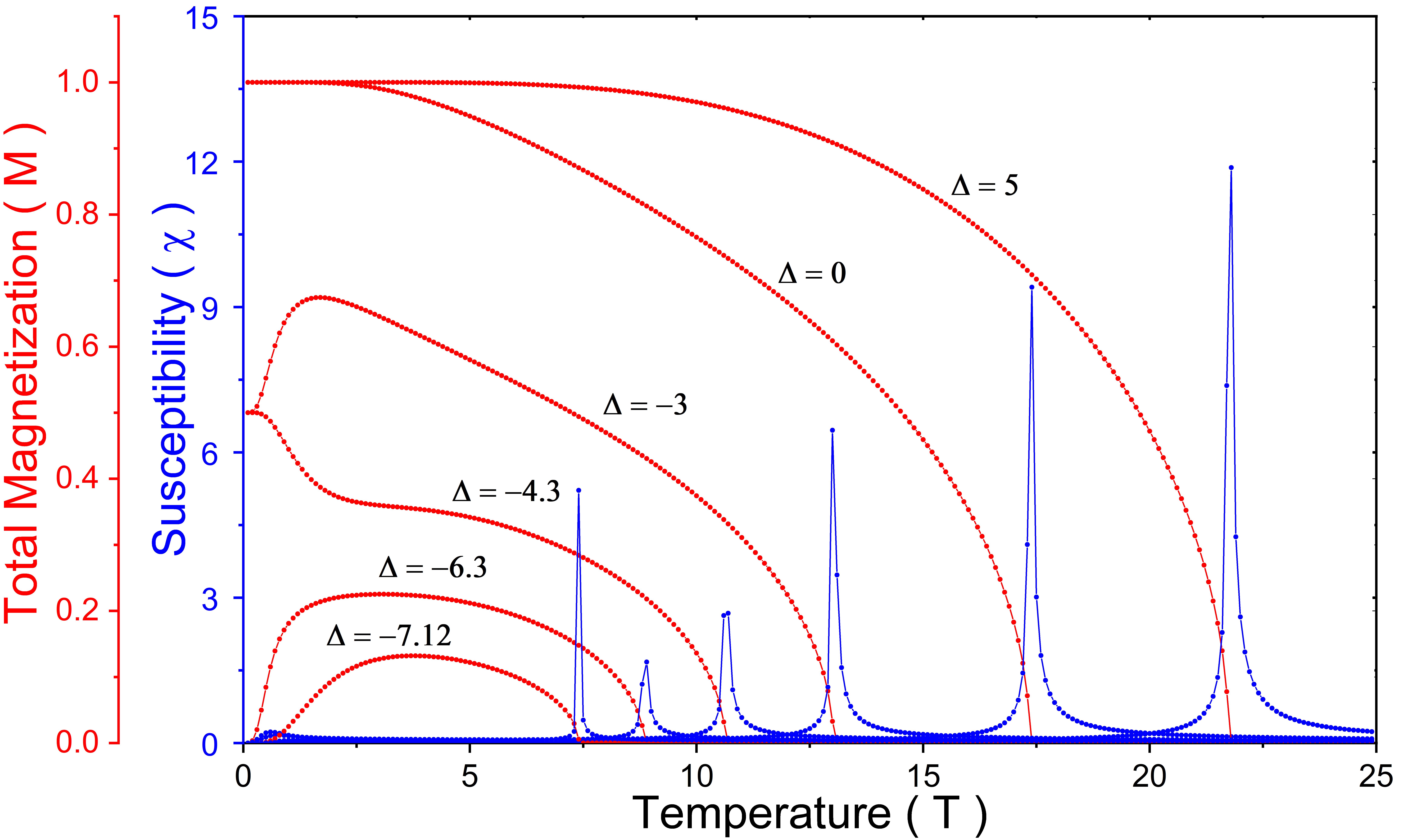}  
\caption{Total magnetization as a function of temperature for selected
values of $\Delta $.}
\label{fig:9}
\end{figure}

To analyze the temperature effect on the total magnetization and the one of
the two sublattices Fe and Mo, we fixed the exchange interaction between two
sublattices ($J_{Fe-Mo} = -3 $) and within each one ($J_{Fe} =
J_{Mo} = 0.1 $) and studied the variation of these magnetizations as
a function of the temperature for different crystalline field ($\Delta $)
values in the absence of the external magnetic field. As the temperature
rises, the order degrades. The variations of the total magnetization
are plotted in \hyperref[fig:9]{Figure \ref{fig:9}} as a function of
temperature for selected values of the crystalline field ($\Delta $). The mixed system depending on the value of the
crystalline field, five types of curves $M_T $ according to the extended Néel  classification \cite{28}: types Q $\Delta=5 $, R $\Delta=0
$, P $\Delta=-3 $, S $\Delta=-4.3 $, and L $\Delta=6.3 $.

\begin{figure}[H]
\centering
\includegraphics[width=0.48\textwidth]{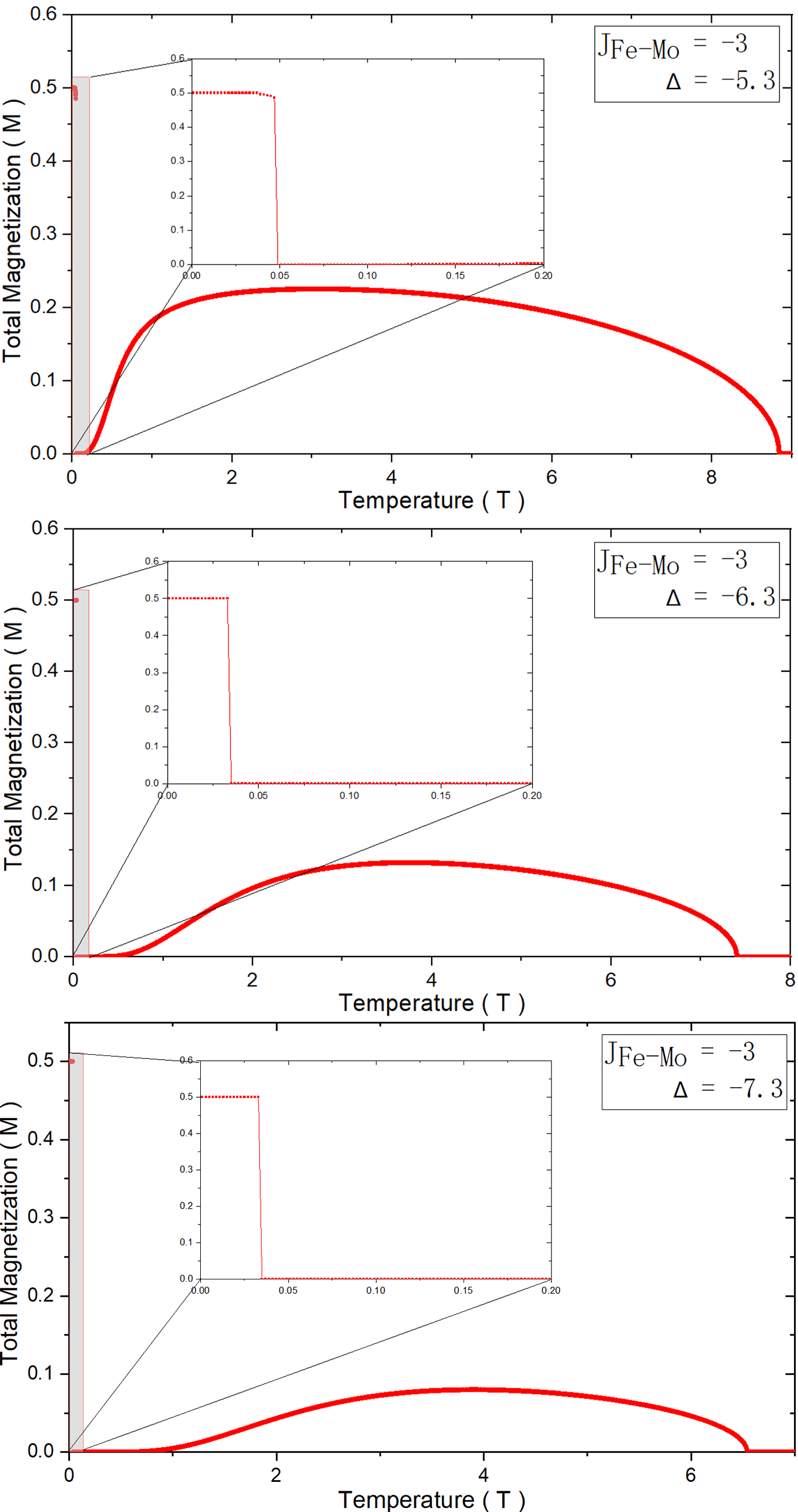}  
\caption{Compensation and Critical Temperatures for Different $\Delta$
Values (-5.3, -6,3, -7.12).}
\label{fig:10}
\end{figure}

For a specific value of the magnetic anisotropy ($\Delta = -5.3 $, $\Delta =
-6.3 $ and$\Delta = -7.3 $), the variation of the total magnetization $M_T $
as a function of the temperature reveals the existence of a point of
compensation. Indeed, the ferrimagnetic interaction between spins can give
rise to zero spontaneous magnetization at a temperature below the critical
temperature. This temperature is called the compensation temperature and is
identified as a compensation point. The total magnetization vanishes at the
compensation temperature $T_{comp} = 0.06 $ before reaching the
critical temperature $T_C = 8.87 $ for $\Delta = -5.3$. For $\Delta = -6.3$, 
$T_C = 7.41 $ and $T_{comp} = 0.031 $. For $\Delta = -7.2$, $T_C =
6.55 $ and $T_{comp} = 0.031 $ \hyperref[fig:10]{Figure \ref{fig:10}}.

\subsection{Hysteresis cycles}

\begin{figure}[H]
\centering
\includegraphics[width=0.48\textwidth]{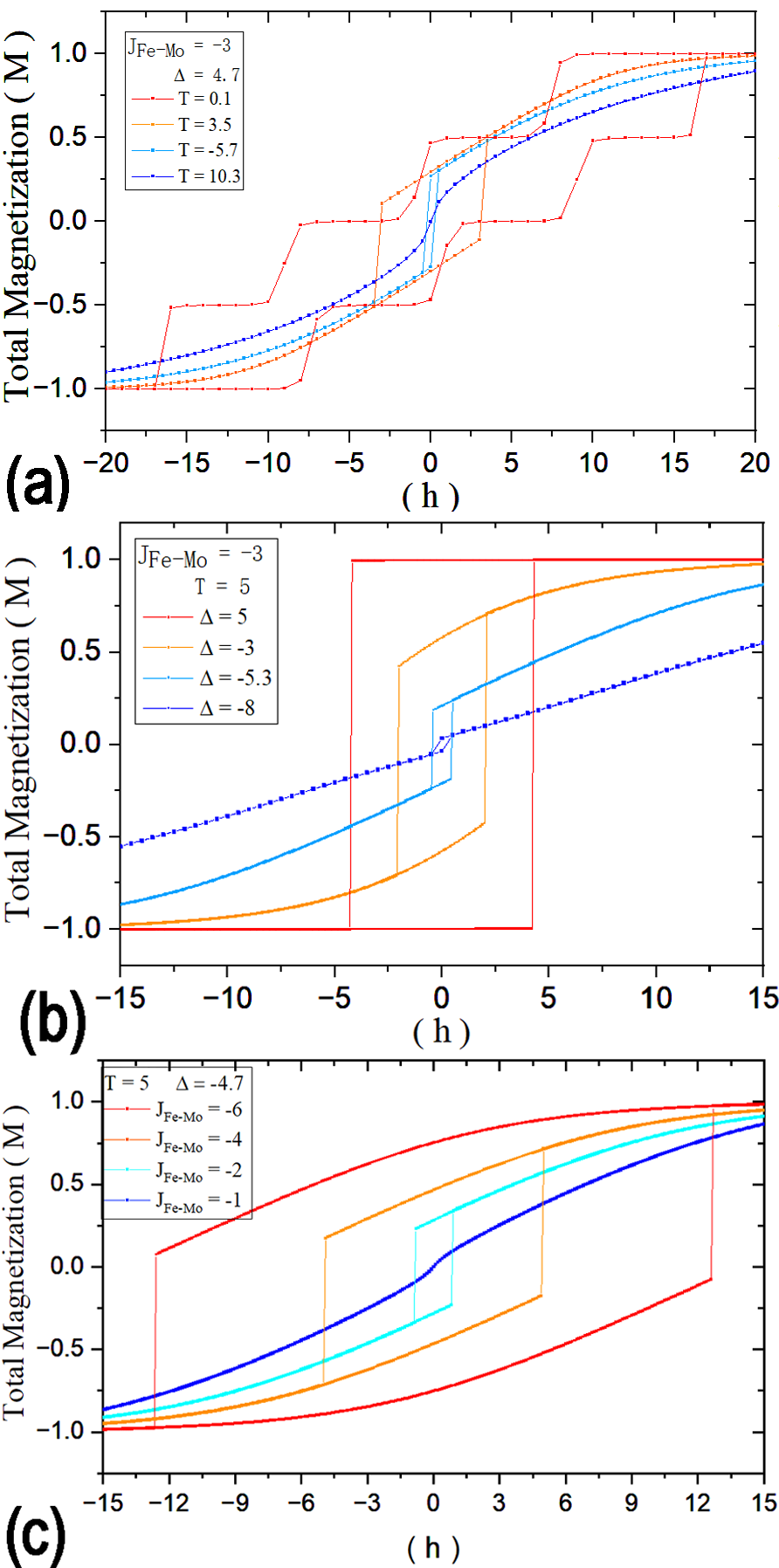}  
\caption{Hysteresis Cycles of the System under Various Conditions
(a) for $J_{Fe-Mo} = -3$, $\Delta = -4.7$, and varying temperatures.
(b) for $J_{Fe-Mo} = -3$, $T = 5$, and varying crystal field.
(c) for variation of $J_{Fe-Mo}$ with $T = 5$ and $\Delta = -4.7$.}
\label{fig:11}
\end{figure}

In order to analyze the magnetic field effects on the system, the hysteresis behaviors was examined to understand the influence of temperature, crystal field ($\Delta$), and exchange interaction ($J_{Fe-Mo}$) on magnetization ($M$) under an external magnetic field ($h$).

we study a series of hysteresis
cycles for different temperature values below and above the critical
temperature. As the temperature increases, the width of the curl decreases
until it disappears as can be seen in \hyperref[fig:11]{Figure \ref{fig:11} (a)}. Below the transition temperature, the cycle presents multiple loops. When
the temperature becomes higher than the transition temperature ($T < T_c = 10.5$),
the width of the single loop is canceled out and thereby hysteresis cycle
disappears. In low temperature (T = 0.5) the high spins can be oriented
parallel to $h$ $\left(\frac{1}{2}, \frac{5}{2}\right)$ and $\left(\frac{1}{2%
}, \frac{3}{2}\right)$ for positive values of $h$, or antiparallel to $h$ $%
\left(-\frac{1}{2}, -\frac{5}{2}\right)$ and $\left(-\frac{1}{2}, -\frac{3}{2%
}\right)$ for negative values of $h$. The other spin states $\left(\frac{1}{2%
}, \frac{1}{2}\right)$, $\left(-\frac{1}{2}, -\frac{1}{2}\right)$, $\left(-%
\frac{1}{2}, \frac{1}{2}\right)$, and $\left(\frac{1}{2}, -\frac{1}{2}\right)
$ are influenced by $h$ and by the crystal field $\Delta$. The disappearance of hysteresis above $T_{c}$ confirms the critical role of temperature in driving phase transitions marking the boundary between ordered ferromagnetic and disordered paramagnetic phases.

Influence crystal field $\Delta$ \hyperref[fig:11]{Figure \ref{fig:11} (b)} the hysteresis cycles show for $\Delta > 0$, sharp, wide loops indicate first-order phase transitions with abrupt magnetization changes. As $\Delta$ decreases to negative values, hysteresis becomes narrower and smoother $\Delta < -3$ , reflecting continuous second-order phase transitions and reduced magnetic stability, demonstrating the importance of the crystal field in controlling energy barriers for magnetization reversal.

In \hyperref[fig:11]{Figure \ref{fig:11} (c)} we see how hysteresis loops depend on the exchange constant $J_{Fe-Mo} $. For strongly negative values
(e.g., $J_{Fe-Mo} = -6 $), the loops show multiple plateaus,
indicating abrupt transitions between stable magnetization states. As $J_{Fe-Mo} $ increases (becoming less negative), the loops become
smoother with fewer plateaus, suggesting a reduction in stable magnetization
states and a shift towards a more continuous magnetization response at
higher $J_{Fe-Mo} $ values.

\section{Monte Carlo simulation}

\begin{figure}[H]
\centering
\includegraphics[width=0.48\textwidth]{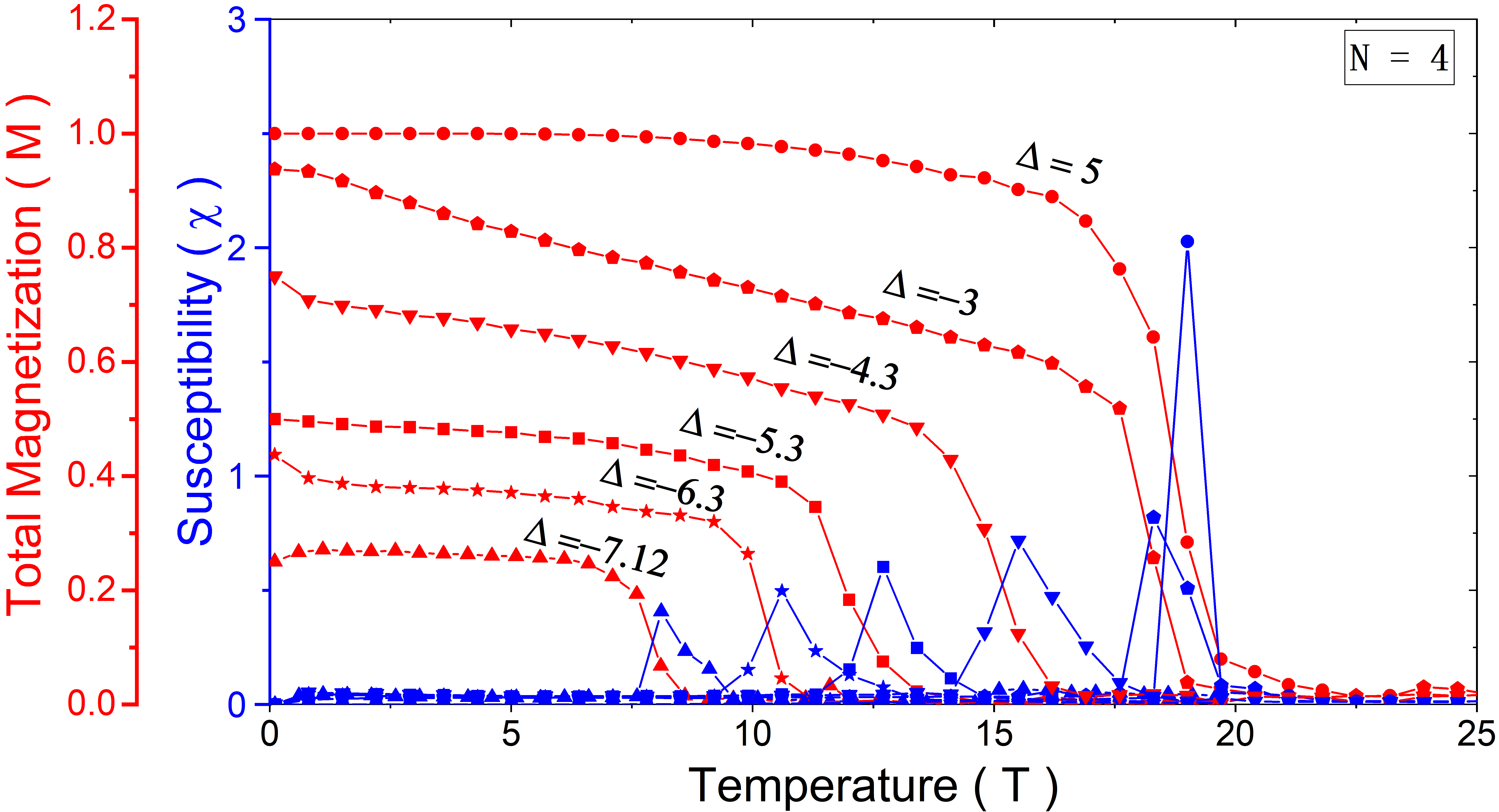}  
\caption{Total magnetization M with variation of T for N=4}
\label{fig15}
\end{figure}

\hyperref[fig15]{Figure \ref{fig15}} show total magnetization $M$ versus
temperature $T$ for a Monte Carlo simulation with $N=4$ and $J_{Fe-Mo
} = -3$, highlighting the effects of various crystal field parameters $\Delta
$. As $\Delta$ becomes more negative, the critical temperature for the
ferromagnetic to paramagnetic transition decreases: $T \approx 20$ for $%
\Delta = 5$, $T \approx 19$ for $\Delta = -3$, $T \approx 16$ for $\Delta =
-4.3$, $T \approx 13$ for $\Delta = -5.3$, and $T \approx 10$ for $\Delta =
-6.3$, $T \approx 8$ for $\Delta = -7.12$. The transitions become sharper
and fluctuations near critical temperatures increase, indicating more
complex phase transition dynamics.

\begin{figure}[]
\centering
\includegraphics[width=0.48\textwidth]{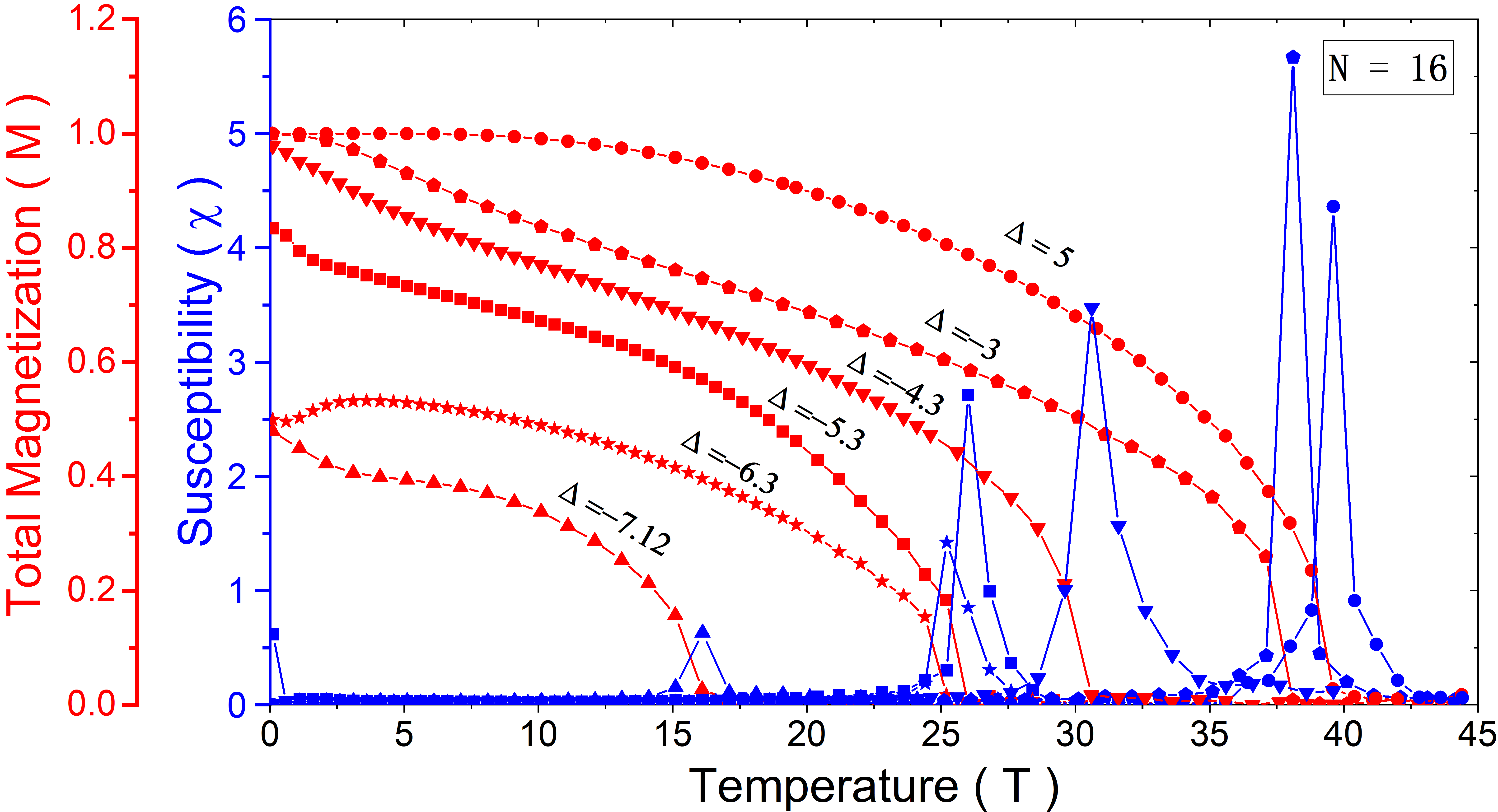}  
\caption{Temperature Dependence of Total Magnetization for N= 16}
\label{fig:16}
\end{figure}

with $J_{Fe-Mo} = -3$ and $N = 16$ reveal how single-ion anisotropy $%
\Delta$ affects temperature-dependent magnetization $M$ \hyperref[fig:16]{%
Figure \ref{fig:16}}. At $\Delta = 5$, magnetization decreases gradually,
and susceptibility peaks broadly at $T \approx 40$, indicating a
second-order transition. For $\Delta = 5$, the critical temperature
increases to $T \approx 40$. As $\Delta$ is reduced to $-3$, $-4.3$, $-5.3$,
and $-6.3$, transitions sharpen, critical temperatures drop, and
susceptibility peaks shift lower, indicating stronger first-order
characteristics. Positive anisotropy stabilizes magnetic order and raises
critical temperatures, while negative anisotropy lowers them.

\section{Conclusion}

In this work, we have studied the magnetic properties in double perovskite
oxide Sr$_{2}$FeMoO$_{6}$ using a mixed spins $\left( \frac{1}{2},\frac{5}{2}%
\right) $ Ising model in a three-dimensional cubic lattice, within the mean
field approximation. Our results are consistent with those obtained by the
standard numerical method of Monte Carlo simulation, confirming the
robustness and reliability of our theoretical approach. We have demonstrated
that the phase diagram of the model shows first-order phase transitions at
low temperatures and second-order transitions at finite temperatures, close
to the critical value. This indicates that the nature of the phase
transition changes with temperature, to help us understanding the thermal
behavior of Sr$_{2}$FeMoO$_{6}$. Our study found that the total
magnetization vanishes at a compensation temperature $T_{comp}=0.06$
with critical temperature $T_{C}=8.87$. For $T_{C}=7.41$, $T_{comp}=0.031$, and for $T_{C}=6.55$, $T_{comp}=0.031$, both due to
ferrimagnetic interactions. This phenomenon is vital for spintronic devices,
where precise magnetization control is essential.

\end{document}